# Calcite precipitation from $CO_2$-$H_2O$-$Ca(OH)_2$ slurry under high pressure of $CO_2$

G. Montes-Hernandez [*a, b], F. Renard [a, c], N. Geoffroy [a], L. Charlet [a], J. Pironon [b]

[a] LGIT, CNRS-OSUG-UJF, Maison de Géosciences, BP 53 X, 38420 Grenoble Cedex 9

[b] G2R, Nancy Université, CNRS, BP 239, 54506 Vandoeuvre lès-Nancy, France

[c] Physics of Geological Processes, University of Oslo, Norway

[*] Corresponding author: German Montes-Hernandez

E-mail address: German.MONTES-HERNANDEZ@obs.ujf-grenoble.fr

german_montes@hotmail.com

**Abstract**

The formation of solid calcium carbonate ($CaCO_3$) from aqueous solutions or slurries containing calcium and carbon dioxide ($CO_2$) is a complex process of considerable importance in the ecological, geochemical and biological areas. Moreover, the demand for powdered $CaCO_3$ has increased considerably recently in various fields of industry. The aim of this study was therefore to synthesize fine particles of calcite with controlled morphology by hydrothermal carbonation of calcium hydroxide at high $CO_2$ pressure (initial $P_{CO2}$=55 bar) and at moderate and high temperature (30 and 90°C). The morphology of precipitated particles was identified by transmission electron microscopy (TEM/EDS) and scanning electron microscopy (SEM/EDS). In addition, an X-ray diffraction analysis was performed to investigate the carbonation efficiency and purity of the solid product.

Carbonation of dispersed calcium hydroxide ($Ca(OH)_{2(s)} + CO_{2(aq)} \rightarrow CaCO_{3(s)} + H_2O$) in the presence of supercritical ($P_T$=90 bar, T=90°C) or gaseous ($P_T$=55 bar, T=30°C) $CO_2$ led to the precipitation of sub-micrometric isolated particles (<1μm) and micrometric agglomerates (<5μm) of calcite. For this study, the carbonation efficiency ($Ca(OH)_2$-$CaCO_3$ conversion) was not significantly affected by PT conditions after 24 h of reaction. In contrast, the initial rate of calcium carbonate precipitation increased from 4.3 mol/h in the "90bar-90°C" system to 15.9 mol/h in the "55bar-30°C" system. The use of high $CO_2$ pressure may therefore be desirable for increasing the production rate of $CaCO_3$, carbonation efficiency and purity, to approximately 48 kg/m$^3$h, 95% and 96.3%, respectively in this study. The dissipated heat for this exothermic reaction was estimated by calorimetry to be -32 kJ/mol in the "90bar-90°C" system and -42 kJ/mol in the "55bar-30°C" system.

## 1. Introduction

Calcium carbonate is an abundant mineral comprising approximately 4% of the Earth's crust. As an inorganic mineral it is widely used both by man for industrial applications and more generally by living organisms during their development. The carbon dioxide balance in the atmosphere (an integral component of the carbon cycle) is partially controlled through the equilibrium cycling between mineralised calcium carbonate and atmospheric carbon dioxide [1]. The formation of solid calcium carbonate ($CaCO_3$) from aqueous solutions or slurries containing calcium and carbon dioxide ($CO_2$) is a complex process of considerable importance in the ecological, geochemical and biological areas. Moreover, the demand for powdered $CaCO_3$ has increased considerably recently in various fields of industry: paper, paint, magnetic recording, textiles, detergents, adhesives, plastics, cosmetics, food, etc. [2].

Calcium carbonate particles have three crystal morphologies, which are generally classified as rhombic calcite, needle-like aragonite and spherical vaterite. Calcite belonging to the Trigonal-Hexagonal-Scalenohedral Class is the most stable phase at room temperature under normal atmospheric conditions, while aragonite and vaterite belong to the Orthorombic-Dipyramidal and Hexagonal-Dihexagonal Dipyramidal Classes, respectively. They are metastable polymorphs that readily transform into the stable calcite phase. The polymorphs of crystalline calcium carbonate particles depend mainly on precipitation conditions such as pH, temperature and supersaturation. This last parameter is considered to be the most important controlling factor [3].

Many experimental studies have reported the synthetic precipitation of the various forms of calcium carbonate. The conditions under which precipitation may be produced, including the importance of initial supersaturation, temperature, pH, hydrodynamics and the presence of impurities and additives, are well known (see for ex. [4-17]). In general, calcite in a supersaturated solution can be precipitated by nucleation in an early stage and subsequent crystal growth in later stages. Nucleation corresponds to the formation of nuclei or critical clusters where crystal growth can occur spontaneously [18-19]. The structure of these nuclei is not known and is too small to be observed directly.

The degree of supersaturation ($S_I$) with respect to calcite is defined as:

$$S_I = \frac{(Ca^{2+})(CO_3^{2-})}{K_{sp}} \qquad (1)$$

where ($Ca^{2+}$) and ($CO_3^{2-}$) are the activities of calcium and carbonate ions in the solution, respectively, and $K_{sp}$ is the thermodynamic solubility product of calcite. It is generally agreed that heterogeneous nucleation can be initiated at a lower degree of supersaturation than homogeneous nucleation and different seed crystals will lower the activation energy barrier, depending on the level of molecular recognition between the seed crystal and the precipitating solid phase [18].

The low viscosity and high diffusivity of $CO_2$ occurring in near-critical or supercritical (SC-) state result in enhanced reaction rates for processes carried out involving these media. Therefore, compressed $CO_2$ has many properties that make it an attractive solvent for industrial chemical processes and/or reactions. SC-$CO_2$ has been used in a number of industrial processes, most notably as an extraction solvent and for pharmaceutical particle engineering [20]. Promising applications in low-cost bulk materials, such as cement, are also being widely investigated in order to improve concrete strength for immobilizing radioactive waste [21, 22]. There has also been a general recent trend in the chemical and petrochemical industry to operate gas absorption units at high pressure for both technical and economic reasons. Recently, the precipitation of calcium carbonate in compressed and supercritical $CO_2$ was proposed as an innovative method for producing calcite by aqueous-carbonation with potential benefits for industrial applications [2, 23]. Note that, for these applications, SC-$CO_2$ is not used as a solvent but as a reagent for calcite precipitation. See for example, the following overall reaction:

$$Ca(OH)_{2(s)} + CO_{2(aq)} \rightarrow CaCO_{3(s)} + H_2O \qquad (2)$$

Finally, the increasing $CO_2$ concentration in the earth's atmosphere, mainly caused by fossil fuel combustion, has led to concerns about global warming. A technology that could possibly contribute to reducing carbon dioxide emissions is the in situ sequestration (geological storage) or ex-situ sequestration (controlled reactors) of $CO_2$ by mineral carbonation, as originally proposed by Seifritz [24] and first studied in detail by Lackner et al. [25]. At the present time, several theoretical and/or experimental studies on $CO_2$ sequestration have been reported in the literature (see for example [26-45]). The basic concept behind mineral $CO_2$ sequestration is to mimic natural weathering processes in which calcium or magnesium silicates are transformed into carbonates:

$$(Ca,Mg)SiO_3(s) + CO_2(g) \rightarrow (Ca,Mg)CO_3(s) + SiO_2(s) \qquad (3)$$

The aim of this study was to synthesize fine particles of calcite with controlled morphology by using hydrothermal carbonation of calcium hydroxide at high $CO_2$ pressure (initial $P_{CO_2}$=55 bar) and at moderate and high temperature (30, 90°C). In addition, a simplified method was proposed to estimate the calcium carbonate production rate, carbonation efficiency, purity and dissipated heat of the exothermic reaction.

Portlandite $Ca(OH)_2$ material was chosen because this mineral is an important component in cement to be used in concrete injection systems for geological storage of $CO_2$. Here, cement carbonation is the main chemical transformation occurring in borehole materials in contact with $CO_2$. This process could be the cause of a possible loss of borehole integrity, inducing leakage of the gas to the surface. In addition, the carbonation of $Ca(OH)_2$ at high $CO_2$ pressure was recently proposed as a novel method to produce fine particles of calcite [2, 23].

## 2. Materials and methods

2   2.1 Carbonation of calcium hydroxide (stirred reactor)

3   One litre of high-purity water with electrical resistivity of 18.2 MΩ cm and 74.1g of

4   commercial calcium hydroxide (provided by Sigma-Aldrich) with 96% chemical purity (3%

5   $CaCO_3$ and 1% other impurities) were placed in a titanium reactor (autoclave with internal

6   volume of two litres). The hydroxide particles were immediately dispersed with mechanical

7   agitation (400 rpm). The dispersion was then heated to 90°C with a heating system adapted to

8   the reactor. When the dispersion temperature was reached, 80.18 g of $CO_2$ (provided by Linde

9   Gas S.A.) were injected in the reactor and the total pressure in the system was immediately

10  adjusted to 90 bar by argon injection (see Fig. 1). Under these T and P conditions, the vapour

11  phase consists mainly of an Ar+$CO_2$ mixture with the $CO_2$ in a supercritical state. In order to

12  evaluate the precipitation (or production) rate, four different reaction durations were

13  considered (0.25, 0.5, 4 and 24 h). The experiments were also carried out at 30°C and 55 bar

14  for reaction durations of 0.25, 4 and 24 h. For this second case, 96.05 g of $CO_2$ were initially

15  injected in the reactor. At 55 bar and 30°C, the vapour phase consists mainly of gaseous $CO_2$.

16  At the end of the experiment, the autoclave was removed from heating system and immersed

17  in cold water. The reaction cell was depressurized during the water cooling period. After

18  water cooling at 35°C (about 15 minutes) the autoclave was disassembled, and the solid

19  product was carefully recovered and separated by centrifugation (30 minutes at 12,000 rpm),

20  decanting the supernatant solutions. Finally, the solid product was dried directly in the

21  centrifugation flasks for 48 h at 60°C and consecutively for 12 h at 110°C in order to

22  eliminate the adsorbed water. The weight of dry solid product was then calculated by the

23  following simple mass balance:

24  $$w_{dry\_product} = w_{flask+dry\_product} - w_{flask} \qquad (4)$$

Note that the reaction cell was depressurized immediately after immersion in cold water. In order to evaluate the effect of the depressurization stage on the final product composition, a test experiment was performed at 90 bar and 90°C for a reaction time of 4 h. In this case, the reactor was depressurized after water cooling at 35°C, for 45 minutes.

2.2 Characterization of solid phase

Morphological analyses of the solid products were performed using a Hitachi 22500 Fevex scanning electron microscope. Isolated fine particles (oriented on carbon Ni-grids) of the starting material and product were also studied using a Jeol 3010 transmission electron microscope, equipped with an energy dispersive X-ray analyser (EDS) to illustrate particle morphology and to identify the precipitated phases. The starting material and solid products were also characterized by X-ray powder diffraction using a Siemens D501 diffractometer with a $\theta$, $2\theta$ geometry. The XRD patterns were recorded in the 5-80° $2\theta$ range using Cok$\alpha$ radiation ($\lambda$=1.7902 Å). The experimental measurement parameters were 8s counting time per 0.02°$2\theta$ step. A Kevex Si(Li) detector is used for this purpose.

**3. Results and discussion**

3.1 Dissipated heat of $Ca(OH)_2$ carbonation

The aqueous carbonation of calcium hydroxide described by the global reaction (2) is an exothermic process which concerns simultaneously the dissolution of calcium hydroxide,

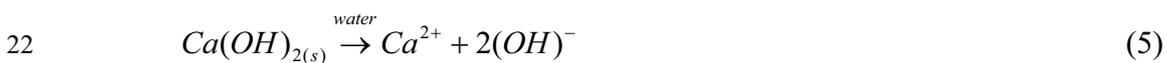

$$Ca(OH)_{2(s)} \xrightarrow{water} Ca^{2+} + 2(OH)^- \qquad (5)$$

and the dissociation of aqueous $CO_2$,

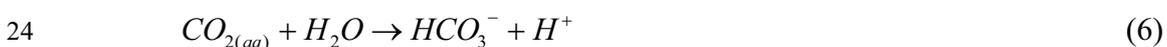

$$CO_{2(aq)} + H_2O \rightarrow HCO_3^- + H^+ \qquad (6)$$

Therefore, the dissipated heat ($q$) for the overall calcium hydroxide carbonation process (reaction (2)) can be calculated by using the calorimetry concept:

$$q = (m)(c)(\Delta T) \tag{7}$$

where $m$ is the mass of water in the reactor (1000 g), $c$ is the specific heat of water (4.19 J/g°C) and $\Delta T$ is the change in water temperature. The temperature change was directly monitored in the reactor as a function of time for each experiment (see for ex. Fig. 2 and Table 1). The average value of $\Delta T$ was estimated to be 7.5°C (±1°C) for the 90bar-90°C system and 10°C (±1°C) for the 55bar-30°C system. This significant variation in $\Delta T$ could be due to the quantity of $CO_2$ dissolved in the aqueous solution because the solubility of $CO_2$ decreases significantly with increase in temperature. In contrast, $CO_2$ solubility increases slightly with increase in pressure (see for ex. [46-47]). Considering the quantity of $Ca(OH)_2$ used in the reaction (74.1 g), the dissipated heats for this exothermic process were -31.4 kJ/mol and -41.9 kJ/mol, respectively.

## 3.2 Carbonation efficiency and purity

The X-ray spectra in Fig. 3 show the qualitative variation in carbonation for the 90bar-90°C and 55bar-30°C systems at different reaction times. Note the high rate of carbonation process and high purity of precipitated calcite. These results suggest a simple mechanism for calcite precipitation, first spontaneous precipitation of amorphous calcium carbonate and then calcite formation. This process can be represented by the following successive chemical reactions:

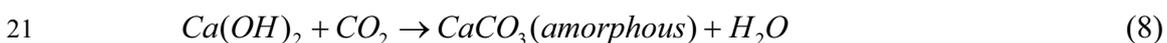
$$Ca(OH)_2 + CO_2 \rightarrow CaCO_3(amorphous) + H_2O \tag{8}$$

and,

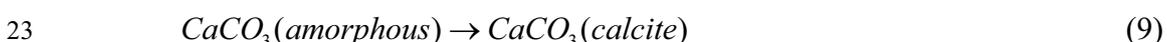
$$CaCO_3(amorphous) \rightarrow CaCO_3(calcite) \tag{9}$$

The metastable crystalline phases of calcium carbonate, such as vaterite and aragonite, were not identified during the calcium hydroxide carbonation process, except in the test experiment, i.e. when the reactor was depressurized after the water cooling stage at 35°C. For this case, crystalline aragonite was also identified by X-ray diffraction (see Fig. 4). Obviously, this crystalline phase was formed during water cooling stage, but the physicochemical mechanism is not clear. The excess $CO_2$ in the system (about 0.9 mol) may play a significant role in this unusual process, possibly with partial dissolution of calcite particles and then aragonite precipitation and calcium hydroxide re-crystallization during reactor depressurization at low temperature.

For this study, the presence of supercritical $CO_2$ did not have a clear effect on the $Ca(OH)_2$ carbonation process. This confirms the low reactivity of molecular $CO_2$ on calcite dissolution [44] and precipitation (this study). In contrast, the temperature of reaction has a significant effect on carbonation rate (see Figure 2). In fact, the precipitation rate is proportional to the quantity of dissolved $CO_2$. This justifies a higher rate of carbonation at low temperature.

From the qualitative characterization by X-ray diffraction discussed above the carbonation efficiency and chemical purity of calcium carbonate were calculated using a mass balance method, based on the theoretical overall carbonation reaction (see Table 2). The carbonation efficiency ($CE$) was then calculated by the following equation:

$$CE = \frac{w_{dry\_product} - w_{Ca(OH)2(initial)}}{w_{theoretical} - w_{Ca(OH)2(initial)}} * 100 \tag{10}$$

where $w_{dry\_product}$ is the experimental mass of solid product [g] (see Eq. 4), $w_{theoretical}$ is the theoretical mass of calcium carbonate calculated by using Eq. (2) [g] and considering 100% of carbonation ($Ca(OH)_2$-$CaCO_3$ transformation), $w_{Ca(OH)2(initial)}$ is the initial mass of calcium hydroxide charged into the reactor [g]. Consequently, the non-reacted calcium hydroxide in the solid product [g] was calculated by:

$$w_{Ca(OH)2(non-reacted)} = w_{Ca(OH)2(initial)} - w_{Ca(OH)2(initial)} * CE \tag{11}$$

Then, the precipitated quantity of calcium carbonate [g] was calculated by a simple mass balance:

$$w_{CaCO3(precipitated)} = w_{dry\_product} - w_{Ca(OH)2(non-reacted)} \qquad (12)$$

Finally, the purity of calcium carbonate was calculated by:

$$Purity = \frac{w_{CaCO3(precipitated)}}{w_{CaCO3(precipitated)} + w_{Ca(OH)2(non-reacted)}} * 100 \qquad (13)$$

All the values are given in Table 2. Note that carbonation efficiency quickly exceeds 90% and reaches a maximum of about 95%. Consequently, the maximum purity value of 96.3% was calculated. The carbonation efficiency and purity of the product obtained by the proposed method could have advantageous technological and environmental applications because it is well-known that the traditional industrial method for carbonation by $CO_2$ bubbling gives lower precipitation efficiencies (see for ex. [2, 23]).

The carbonation of dispersed calcium hydroxide (Eq. 2) in presence of supercritical or gaseous $CO_2$ led to the precipitation of sub-micrometric isolated particles (<1 μm) and micrometric agglomerates (<5 μm) of calcite (see Figure 5).

3.3 Fitting of kinetic experimental data

Several kinetic models including first-order, pseudo-first-order, second-order, pseudo-second-order, parabolic diffusion and power function kinetic expressions are reported in the literature for fitting the kinetic experimental data of a solid-liquid separation process. For this study, the kinetic experimental data concern calcium carbonate precipitation; in this case, the best fit (attested by a correlation factor close to 1) of the experimental data was achieved when using a pseudo-second-order kinetic model according to the following expression:

$$\frac{d[Mol_{CaCO3,t}]}{dt} = k_p \left( Mol_{CaCO3,max} - Mol_{CaCO3,t} \right)^2 \qquad (14)$$

1  where $k_p$ is the rate constant of calcite precipitation [1/mol h] for a given initial dose of

2  $Ca(OH)_2$, $Mol_{CaCO3,max}$ is the maximum precipitated quantity of calcium carbonate at

3  equilibrium [mol], $Mol_{CaCO3,t}$ is the precipitated quantity of calcium carbonate at any time, $t$,

4  [mol].

5  The integrated form of Equation (14) for the boundary conditions $t = 0$ to $t = t$ and $Mol_{CaCO3,t}$

6  $= 0$ to $Mol_{Caco3,t} = Mol_{CaCO3,t}$, is represented by a hyperbolic equation:

7
$$Mol_{CaCO3,t} = \frac{Mol_{CaCO3,\max} * t}{\left(\dfrac{1}{k_P * Mol_{CaCO3,\max}}\right) + t} \tag{15}$$

8  In order to simplify the experimental data, fitting a novel constant can be defined

9  "$(1/k_p*Mol_{CaCO3,max}) = t_{1/2}$". Physically, this constant represents the time after which half of the

10  maximum precipitated quantity of calcium carbonate was obtained. In the current study, $t_{1/2}$ is

11  called "half-precipitation time" and can be used to calculate the initial rate of calcium

12  carbonate precipitation, $v_{0,p}$, [mol/h] by the following expression:

13
$$v_{0,p} = \frac{Mol_{CaCO3,\max}}{t_{1/2}} = k_p \left(Mol_{CaCO3,\max}\right)^2 \tag{16}$$

14  The fitting of experimental kinetic curves ($Mol_{CaCO3,t}$ vs. $t$) using Equation (15) is shown in

15  Figure 6. The parameters $t_{1/2}$ and $Mol_{CaCO3,max}$ were estimated by applying non-linear

16  regression by the least squares method.

17  Finally, based on the initial rate of calcium carbonate precipitation the optimum production

18  rate, $P_{rate}$ [kg/m³h] of calcium carbonate can be calculated for a given dose of calcium

19  hydroxide using the following expression:

20
$$P_{rate} = \frac{Mol_{CaCO3,\max} * M_{CaCO3}}{t_{1/2} * V_{reactor}} \tag{17}$$

21  where $M_{CaCO3}$ is the molar mass of calcium carbonate and $V_{reactor}$ is the effective volume of

22  the reactor, (2L) for this study.

The calculations using Eq. (17) show that the calcium carbonate production rate in the 55bar-30°C system (798.6 kg/m$^3$h, $v_{0,p}$=15.9 mol/h) is higher than the production rate in the 90bar-90°C system (213.17 kg/m$^3$h, $v_{0,p}$=4.3 mol/h). This confirms the qualitative observations discussed above. It is important to note that the initial rate of calcium carbonate precipitation, and consequently the production rate calculation, depends only on the $t_{1/2}$ parameter, which in turn depends on the PT conditions.

Note that the production rate was calculated using the $t_{1/2}$ parameter. Obviously, this is an ideal calculation. A more realistic calculation could be suggested based on the experiments and solid product characterization using a 1h reaction time for both the systems studied in order to obtain better product maturity. In this case, the average production rate is estimated to be 47.94 kg/m$^3$h.

**4. Conclusion**

In conclusion, the results of this research show that the carbonation of dispersed calcium hydroxide in water with co-existence of supercritical or gaseous $CO_2$ leads to the precipitation of sub-micrometric isolated particles (<1 µm) and micrometric agglomerates (<5 µm) of calcite. In this study, the carbonation efficiency (Ca(OH)$_2$-CaCO$_3$ conversion) is not significantly affected by PT conditions after 24 h of reaction. In contrast, the initial rate of calcium carbonate precipitation increases from 4.3 mol/h in the "90bar-90°C" system to 15.9 mol/h in the "55bar-30°C" system. The use of high $CO_2$ pressure (initial $P_{CO2}$=55 bar) could represent an improvment for increasing the CaCO$_3$ production rate, carbonation efficiency and purity, which were respectively equal to 48 kg/m$^3$h, 95% and 96.3% in this study. The dissipated heat for this exothermic reaction was estimated by calorimetry concept, -32 kJ/mol for the "90bar-90°C" system and -42 kJ/mol for the "55bar-30°C" system. The results presented here suggest that the carbonation of calcium hydroxide in presence of supercritical or gaseous $CO_2$ could be a powerful technique for producing fine-particles of calcite on an industrial scale. This research also has important ecological implications for the ex-situ mineral sequestration of $CO_2$ by alkaline liquid-solid waste (ex. fly ash, bottom ash, Ca/Mg-rich silicates, alkaline waste water, etc.).

**Acknowledgements**

The authors are grateful to the National Research Agency, ANR (GeoCarbone-CARBONATATION and INTEGRITY projects) and the National Research Council (CNRS), France, for providing a financial support for this work. This study has also been financed through collaboration between the University of Grenoble (German Montes-Hernandez, François Renard) and Gaz de France (Christophe Rigollet, Samuel Saysset, Rémi Dreux).

## References

[1] Dickinson, S. R.; Henderson, G. E.; McGrath, K. M., J. Crystal Growth 244 (2002) 369.

[2] Domingo, C.; Loste, E.; Gomez-Morales, J.; Garcia-Carmona, J.; Fraile, J., The Journal of Supercritical Fluids 36 (2006) 202.

[3] Han, Y. S.; Hadiko, G.; Fuji, M.; Takahashi, M., J. Crystal Growth 276 (2005) 541.

[4] Moore, L.; Hopwood, J D.; Davey, R. J., J. Crystal Growth 261 (2004) 93.

[5] Westin, K. J., Rasmuson, A. C., Journal of Colloids and Interface Science 282 (2005) 370.

[6] Tsuno, H., Kagi, H., Akagi T., Bull. Chem. Soc. Jpn. 74 (2001) 479.

[7] Fujita, Y., Redden, G. D., Ingram, J., Cortez, M. M., Ferris, G., Smith, R. W., Geochimica et Cosmochimica Acta 68/15 (2004) 3261.

[8] Freij, S. J., Godelitsas, A., Putnis, A., J. Crystal Growth 273 (2005) 535.

[9] Gower, L. A., Tirrell, D. A., J. Crystal Growth 191 (1998) 153.

[10] Jonasson, R. G., Rispler, K., Wiwchar, B., Gunter, W. D., Chemical Geology 132 (1996) 215.

[11] Chrissanthopoulos, A., Tzanetos, N. P., Andreopoulou, A. K., Kallitsis, J., Dalas, E., J. Crystal Growth 280 (2005) 594.

[12] Menadakis, M., Maroulis, G., Koutsoukos, P. G., Computational Materials Science 38 (2007) 522.

[13] Dousi, E., Kallitsis, J., Chrisssanthopoulos, A., Mangood, A. H., Dalas, E., J. Crystal Growth 253 (2003) 496.

[14] Pastero, L., Costa, E., Alessandria, B., Rubbo, M., Aquilano, D., J. Crystal Growth 247 (2003) 472.

[15] Lee, Y. J., Reeder, R., Geochemica et Cosmochemica Acta 70 (2006) 2253.

[16] Temmam, M., Paquette, J., Vali, H., Geochemica et Cosmochemica Acta 64/14 (2000) 2417.

[17] Dalas, E., Chalias, A., Gatos, D., Barlos, K., Journal of Colloids and Interface Science 300 (2006) 536.

[18] Stumm, W. and Morgan J. J., Aquatic Chemistry (1996) Wiley-Interscience

[19] Lin, Y. P. and Singer P. C., Geochimica et Cosmochimica Acta 69/18 (2005) 4495.

[20] McHugh, M. and Krukonis V. Supercritical Fluid Extraction: Principles and Practices (1994) Butterworth-Heineman, US.

[21] Jones, R. H. (1999) US Patent 5, 965, 201.

[22] Ginneken; L. V.; Dutré, V. Adriansens W., Weyten H., The Journal of Supercritical Fluids 30/02 (2004)175.

[23] Domingo, C.; Garcia-Carmona, J.; Loste, E.; Fanovich, A.; Fraile, J. Gomez-Morales J., J. Crystal Growth 271 (2004) 268.

[24] Seifritz, W., Nature 345 (1990) 486.

[25] Lackner, K. S.; Wendt, C. H.; Butt, D. P.; Joyce, E. L.; Sharp, D. H., Energy 20/11 (1995) 1153.

[26] Knauss, K. G.; Johnson, J. W.; Steefel, C. I., Chemical Geology 217 (2005) 339.

[27] Pruess, K. T.; Xu, J.; Apps, J.; García, SPE Journal (2003) 49.

[28] Bachu, S., Energy Convers Mgmt. 43 (2002) 87.

[29] Huijgen, W. J. J.; Witkamp, G-J.; Comans R. N. J., Environmental Science and Technology (2005) 9676.

[30] Rosenbauer, R. J.; Koksalan, T.; Palandri J. L., Fuel Processing Technology 86 (2005) 1581.

[31] Kaszuba, J. P.; Janecky, D. R.; Snow, M. G., Chemical Geology 217 (2005) 277.

[32] Kaszuba, J. P.; Janecky, D. R.; Snow, M. G., Applied Geochemistry 18 (2003) 1065.

[33] Palandri, J. L.; Rosenbauer, R. J.; Kharaka, Y. K., Applied Geochemistry 20 (2005) 2038.

[34] Giammar, D. E.; Bruant Jr., R. G.; Poters, C. A., Chemical Geology 217 (2005) 257.

[35] Gunter, W. D.; Perkins, E. H.; Hutcheon I., Applied Geochemistry 15 (2000) 1085.

[36] Okamoto, I.; Li, X.; Ohsumi, T., Energy 30 (2005) 2344.

[37] Park, A-H. A.; Fan, L-S., Chemical Engineering Science 59 (2004) 5241.

[38] Rendek, E.; Ducom, G.; Germain, P., Journal of Hazardous Materials B128 (2006) 73.

[39] Meima, J. A.; Van der Weijden, D.; Eighmy, T. T.; Comans, R. N. J., Applied Geochemistry 17 (2002) 1503.

[40] Regnault, O.; Lagneau, V.; Catalette, H.; Schneider, H., Comptes Rendus Geoscience. (2005) 1331.

[41] Xu, T.; Apps, J. A.; Pruess, K., Chemical Geology 217 (2005) 295.

[42] Bachu, S.; Adams, J. J., Energy Conversion and Management 44 (2003) 3151.

[43] White, S. P.; Allis, R. G.; Moore, J.; Chidsey, T.; Morgan, C.; Gwynn, W.; Adams, M., Chemical Geology 217 (2005) 387.

[44] Pokrosvky, O. S., Gobulev, S. V., Schott, J., Chemical Geology 217 (2005) 239.

[45] Le Guen, Y., Renard, F., Hellmann, R., Collombet, M., Tisserand, D., Brosse, E., Gratier, J. P., Journal of Geophysical Research 112 (2007) B05421.

[46] Duan, Z., Sun, R., Chemical Geology 193 (2003) 257.

[47] Duan, Z., Sun, R., Zhu, C., Chou, I.-M., Mar. Chem. 98 (2006) 131.

1   Table 1. Average $\Delta T$ calculation for "90bar-90°C" and "55bar-30°C" systems

| Experiment | $T_{initial}$ | $T_{maximal}$ | $\Delta T = T_{maximal} - T_{initial}$ |
|---|---|---|---|
| 90bar-90°C system | | | |
| 0.25h | 90 | 97 | 7 |
| 0.50h | 90 | 98 | 8 |
| 4h | 90 | 98 | 8 |
| 24h | 90 | 97 | 7 |
| | | | $\overline{\Delta T} = 7.5$ |
| 55bar-30°C system | | | |
| 0.25h | 30 | 40 | 10 |
| 4h | 30 | 41 | 11 |
| 24h | 30 | 40 | 10 |
| | | | $\overline{\Delta T} = 10.33 \cong 10$ |

2   $T_{initial}$ is the initial temperature before the carbonation reaction
3   $T_{maximal}$ is the maximum temperature reached during the carbonation reaction
4   $\Delta T$ is the change in water temperature during the carbonation reaction
5

Table 2. Carbonation efficiency (*CE*) and chemical purity calculations for 90bar-90°C and 55bar-30°C systems using a mass balance method.

| Time, [h] | $w_{dry\_product}$ [g] | *CE* [%] | $w_{CaCO3(precipitated)}$ [g] | $w_{Ca(OH)2(non-reacted)}$ [g] | Purity[*] [%] |
|---|---|---|---|---|---|
| | | 90bar-90°C system | | | |
| 0.25 | 85.34 | 43.3 | 43.3 | 42.0 | 50.7 |
| 0.50 | 93.42 | 74.3 | 74.4 | 19.0 | 79.7 |
| 4 | 97.91 | 91.6 | 91.7 | 6.2 | 93.7 |
| 24 | 98.04 | 92.1 | 92.2 | 5.8 | 94.0 |
| | | 55bar-30°C system | | | |
| 0.25 | 94.22 | 77.4 | 77.5 | 16.7 | 82.3 |
| 4 | 98.80 | 95.0 | 95.1 | 3.6 | 96.3 |
| 24 | 98.75 | 94.9 | 94.9 | 3.8 | 96.2 |

[*] Taking into account only two solid phases ($CaCO_3$ and $Ca(OH)_2$)

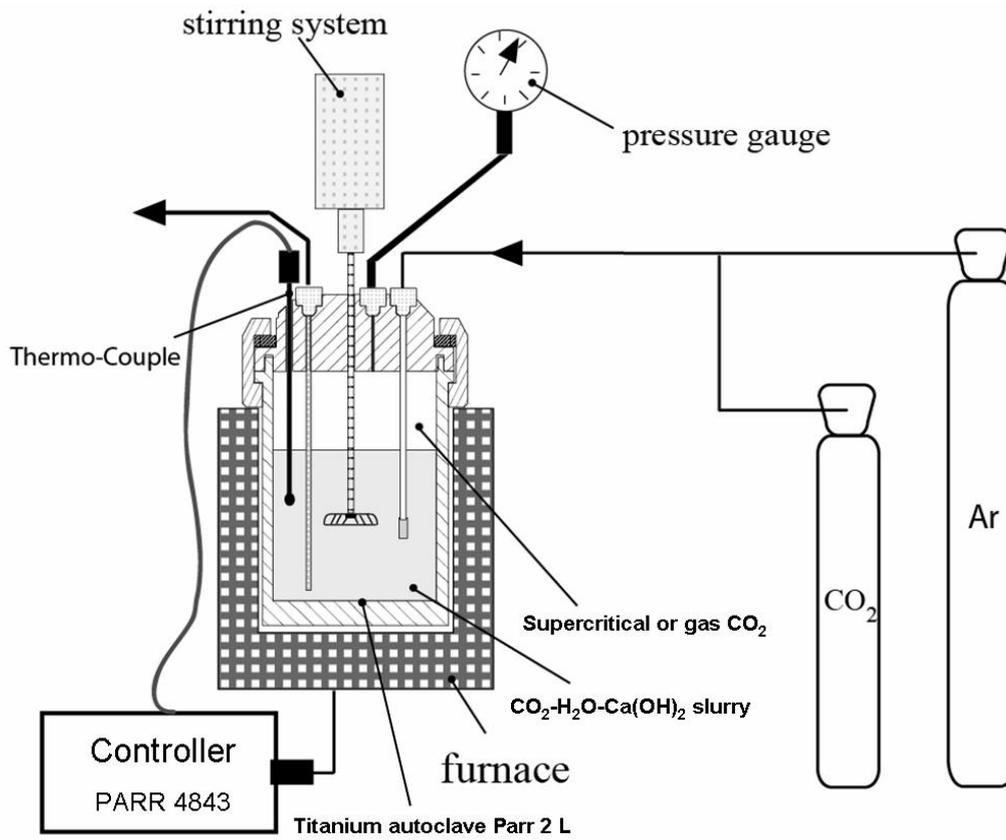

2     Figure 1. Schematic experimental system for calcite precipitation from $CO_2$-$H_2O$-Ca(OH)$_2$

3        slurry in presence of supercritical (90 bar and 90°C) and gaseous (55 bar and 30°C) $CO_2$.

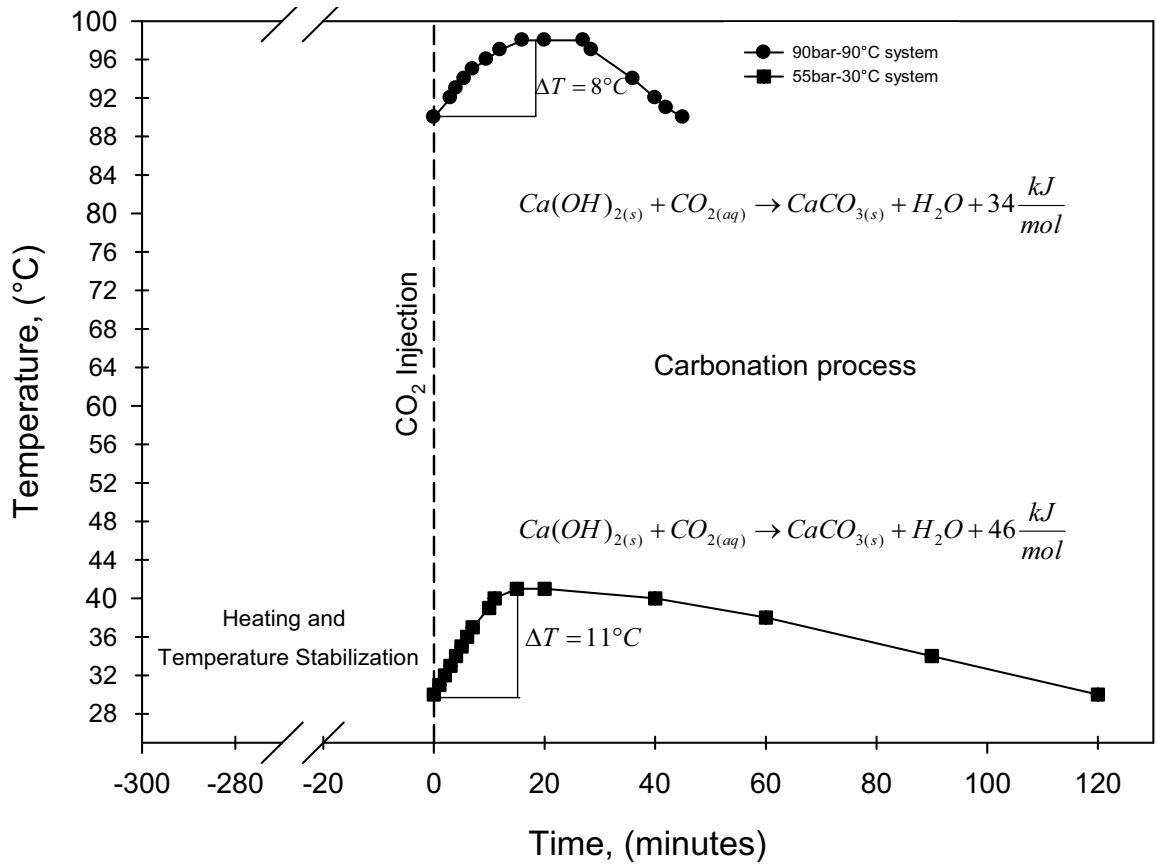

Figure 2. Temperature variation during calcite precipitation for "90bar-90°C" and "55bar-30°C" systems. Estimation of ΔT (temperature change) and calculation of dissipated heat.

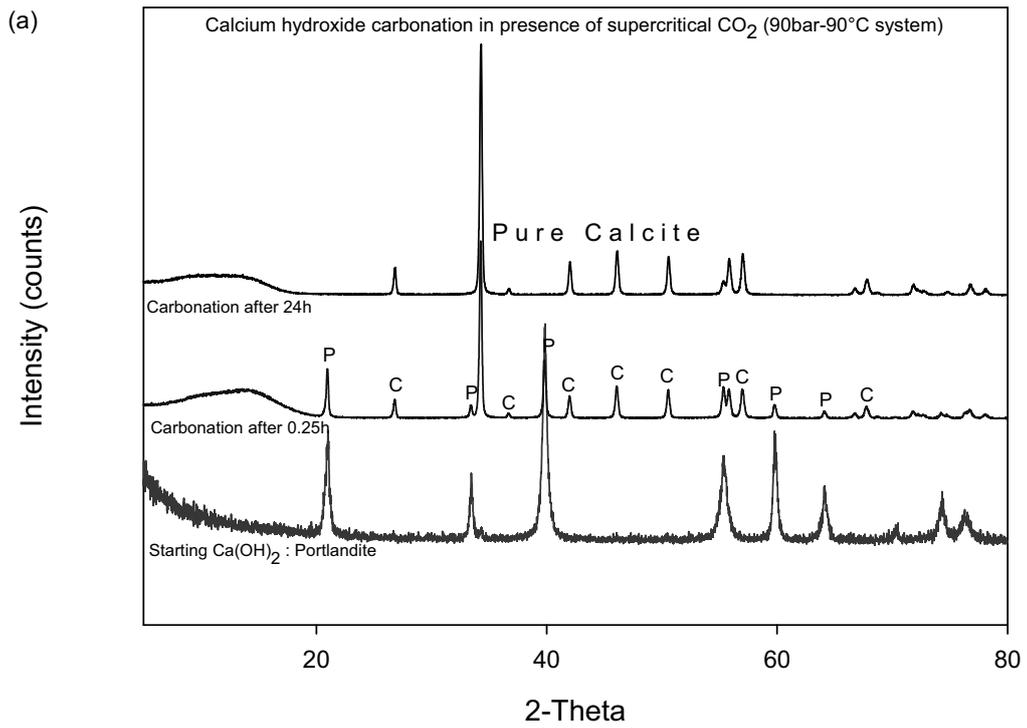

1    a)

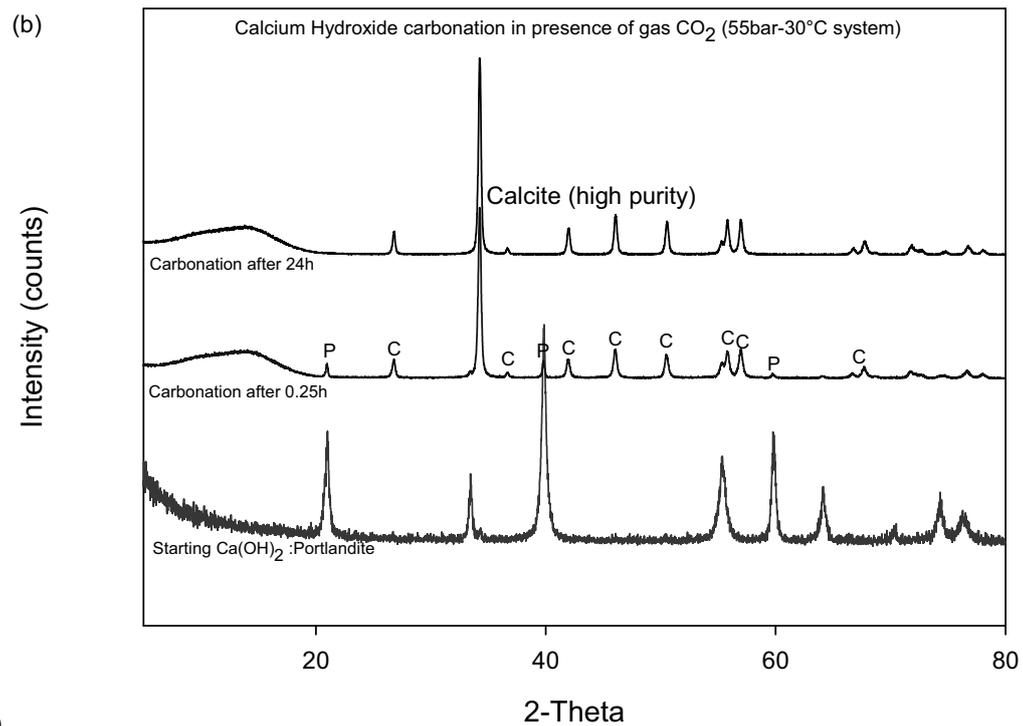

2    b)

3    Figure 3. XRD measurements of initial calcium hydroxide and solid products. (a) Ca(OH)$_2$

4    carbonation in presence of supercritical CO$_2$ and (b) gaseous CO$_2$ at different reaction

5    times of. P: Portlandite, C: Calcite.

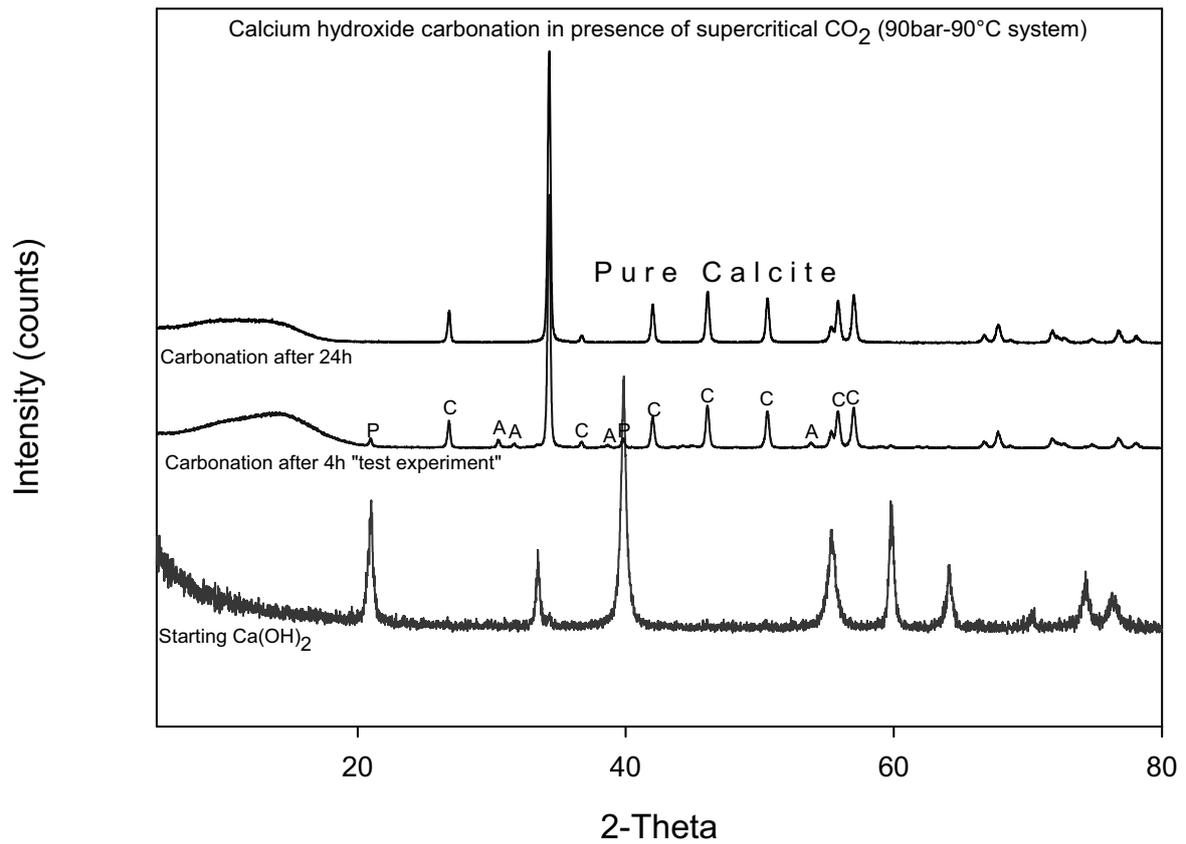

Figure 4. Test experiment "depressurization stage of reactor after water cooling at 35°C". Formation of the crystalline aragonite during the water cooling stage. P: Portlandite, C: Calcite, A:Aragonite.

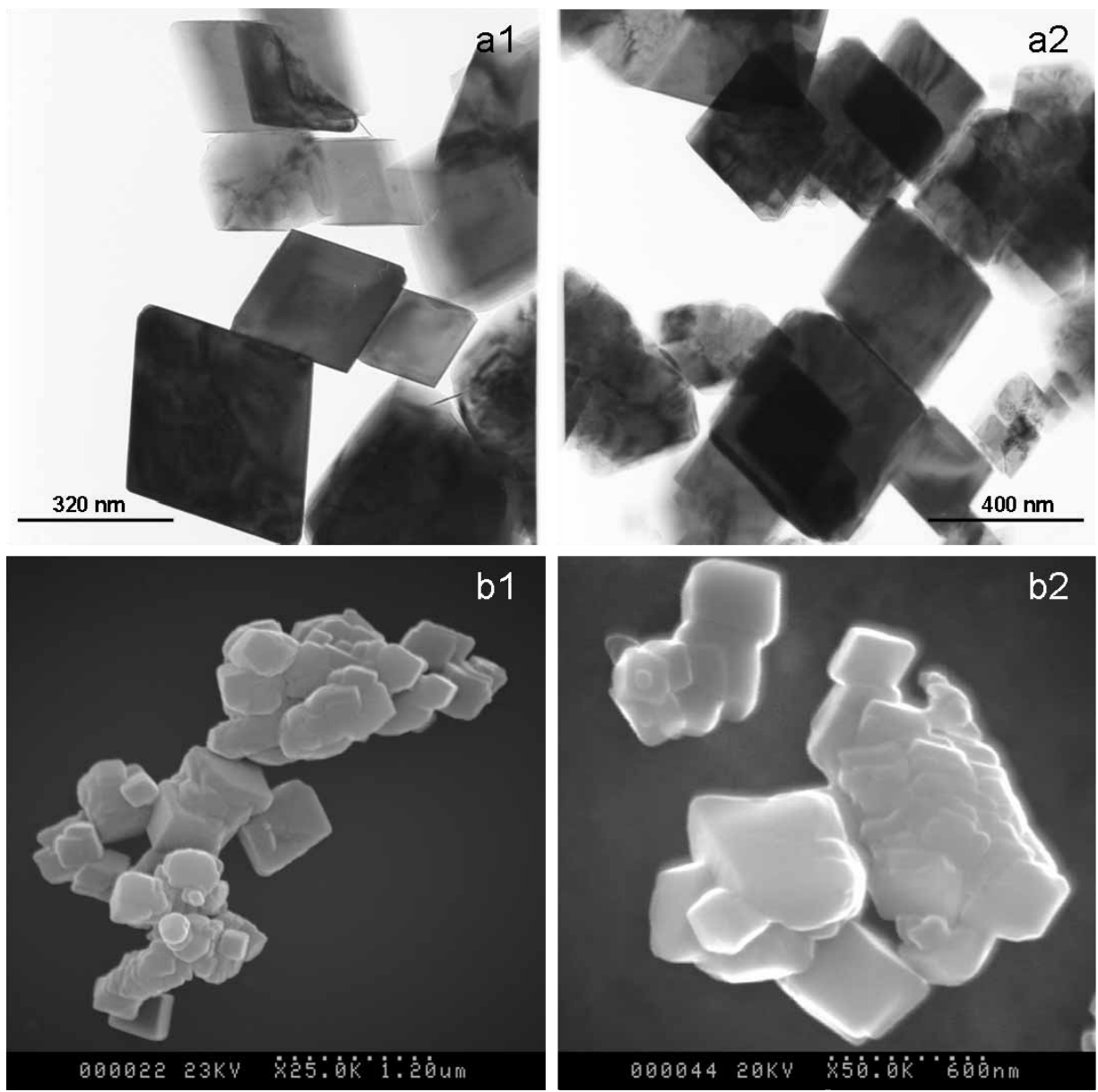

Figure 5. Calcite particles precipitated from $CO_2$-$H_2O$-$Ca(OH)_2$ slurry in presence of supercritical $CO_2$. TEM micrographs at two different magnifications (a1, a2); SEM micrographs at two different magnifications (b1, b2).

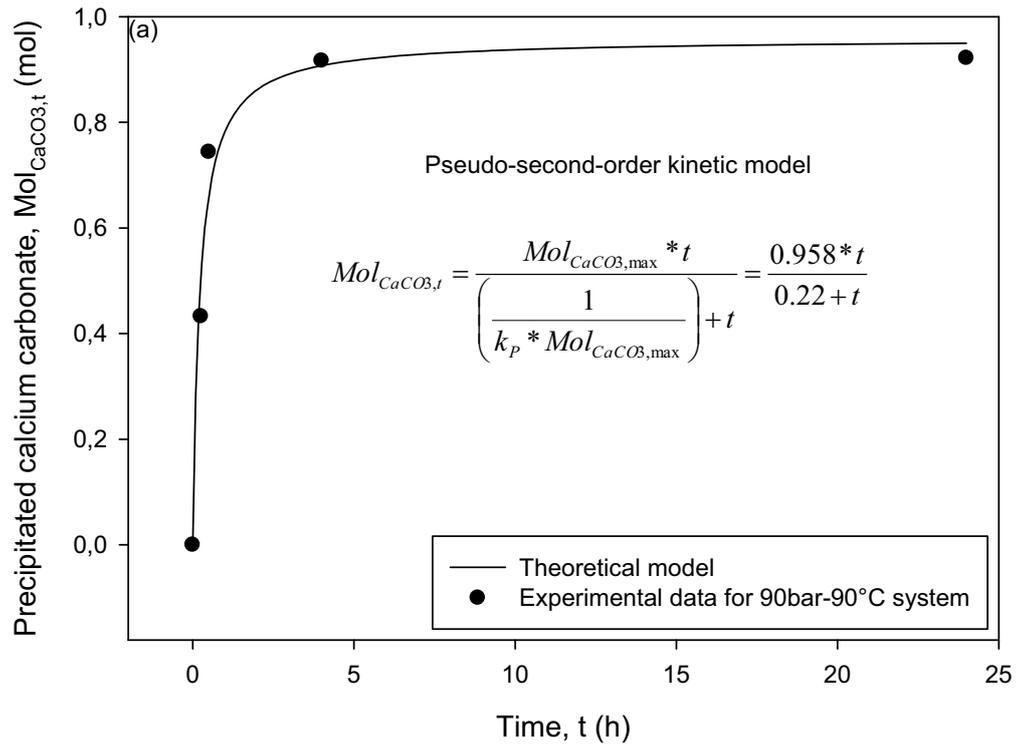

1    a)

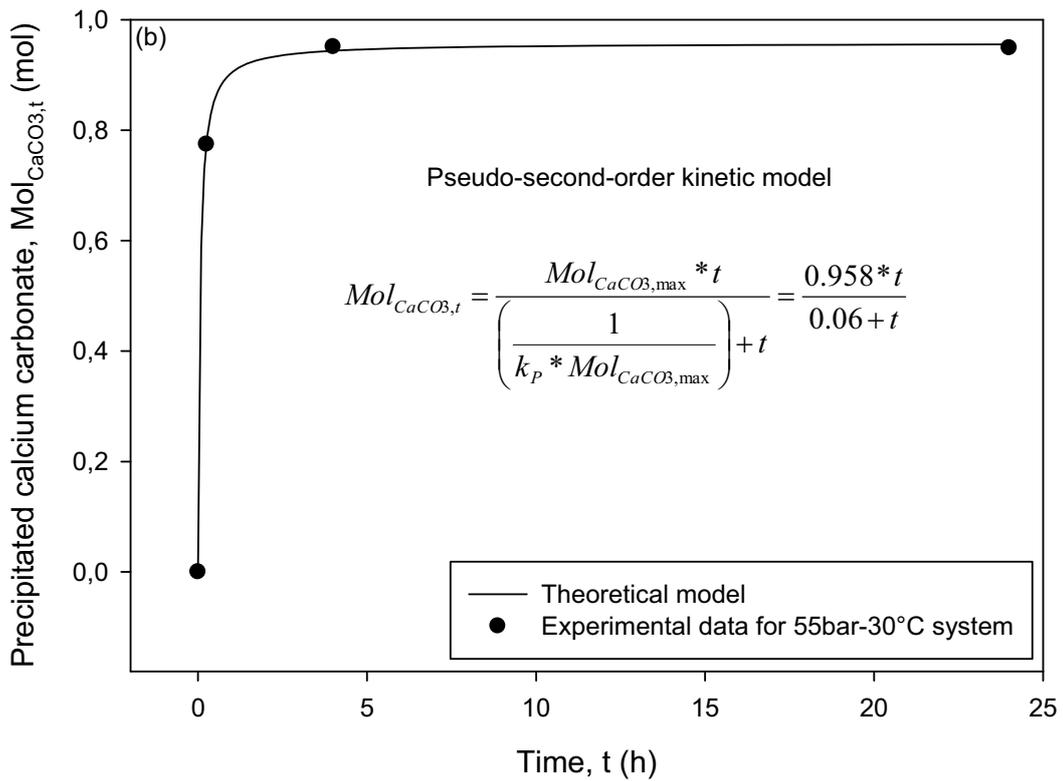

2    b)

3    Figure 6. Fitting of kinetic experimental data on the calcium carbonate precipitation (a) for

4    "90bar-90°C" and (b) "55bar-30°C" systems.